\pdfoutput=1 
\documentclass{JINST}

\usepackage[caption=false]{subfig}

\usepackage{array}
\newcolumntype{x}[1]{>{\centering\let\newline\\\arraybackslash\hspace{0pt}}p{#1}}

\def\papertitle{Measurement of Radiation Damage of Water-based Liquid Scintillator and Liquid Scintillator}
\title{\papertitle}

\author{L. J. Bignell$^a$\thanks{Corresponding author.}~,
        M. V. Diwan$^a$,
        S. Hans$^b$,
        D. E. Jaffe$^a$,
        R. Rosero$^b$,
        S. Vigdor$^c$,
        B. Viren$^a$,
        E. Worcester$^a$,
        M. Yeh$^b$,
        and C. Zhang$^a$\\
\llap{$^a$}Physics Department, Brookhaven National Laboratory,\\ Upton NY, USA\\
\llap{$^b$}Chemistry Department, Brookhaven National Laboratory,\\ Upton NY, USA\\
\llap{$^c$}Phenix Medical LLC, Bloomington, \\IN 47404, USA\\
E-mail: \email{lbignell@bnl.gov}}

\abstract{
Liquid scintillating phantoms have been proposed as a means to perform real-time 3D dosimetry for proton therapy treatment plan verification. We have studied what effect radiation damage to the scintillator will have upon this application. We have performed measurements of the degradation of the light yield and optical attenuation length of liquid scintillator and water-based liquid scintillator after irradiation by 201 MeV proton beams that deposited doses of approximately 52 Gy, 300 Gy, and 800 Gy in the scintillator. Liquid scintillator and water-based liquid scintillator (composed of 5\% scintillating phase) exhibit light yield reductions of $1.74 \pm 0.55 \%$ and $1.31 \pm 0.59 \%$ after $\approx$ 800 Gy of proton dose, respectively. Whilst some increased optical attenuation was observed in the irradiated samples, the measured reduction to the light yield is also due to damage to the scintillation light production. Based on our results and conservative estimates of the expected dose in a clinical context, a scintillating phantom used for proton therapy treatment plan verification would exhibit a systematic light yield reduction of approximately 0.1\% after a year of operation.
}


\begin{document}

\maketitle

\section{Introduction and Purpose}

Dose verification Quality Assurance (QA) measurements are critical in proton therapy for the validation of patient treatment plans \cite{ICRU78ch6}. These measurements are especially important in Intensity Modulated Proton Therapy (IMPT), which scans a focused proton beam spot over the target region many times at different energies and intensities to improve dose conformity. IMPT carries increased treatment planning complexity relative to passively scattered treatment modalities, making the need for dose verification over all beam energy layers especially critical.

An emerging concept for proton therapy dose verification systems is the use of liquid scintillator phantoms with position sensitive photon sensors to image the 3D dose distribution with the phantom volume. These novel detectors present several advantages over the current QA measurement systems that rely upon the physical movement of a two-dimensional array of gaseous detectors within a water phantom \cite{Beddar2009etal, Kroll2013etal}. In particular, liquid scintillator-based systems permit sub-millimeter spatial resolution \cite{Archambault2012etal}, and have the potential to perform validation measurements over all energy layers in no more time than is required for the patient's exposure to the beam. This real-time dose verification is an especially attractive feature of the active phantom approach, as it facilitates a high patient throughput for clinical operations.

Water-based Liquid Scintillator (WbLS) is a newly-developed water-scintillator emulsion \cite{Yeh2011etal} that is a promising candidate material for next-generation particle physics detectors \cite{Alonzo2014etal}. 
As WbLS is predominantly water, it offers improved water-equivalence for a scintillator-filled dose-verification phantom, thereby offering superior dosimetry performance relative to traditional liquid scintillator. WbLS also largely mitigates the flammability and toxicity hazards that are traditionally associated with liquid scintillators. We are currently exploring the possibility of using WbLS for dose verification QA in proton therapy.

One aspect of the use of scintillating materials for dose-verification systems that has yet to be comprehensively investigated is the potential for radiation damage to affect the scintillating performance of the phantom volume. Such damage could lead to systematic measurement errors due to a variable scintillation light yield as a function of dose deposited. In this study, we have quantified the relationship between accumulated dose and scintillating light yield for pure liquid scintillator and WbLS samples. Our measurements of changes to the light yield have been assessed in terms of the 
expected effects upon QA detector operation.

\setlength{\tabcolsep}{0pt}
\begin{figure*}[!t]
\centering
\begin{tabular}{cc}
    \subfloat[ ] {\label{subfig:BeamSetup} \includegraphics[trim={2.5cm 0 2cm 0},clip,width=0.49\textwidth]{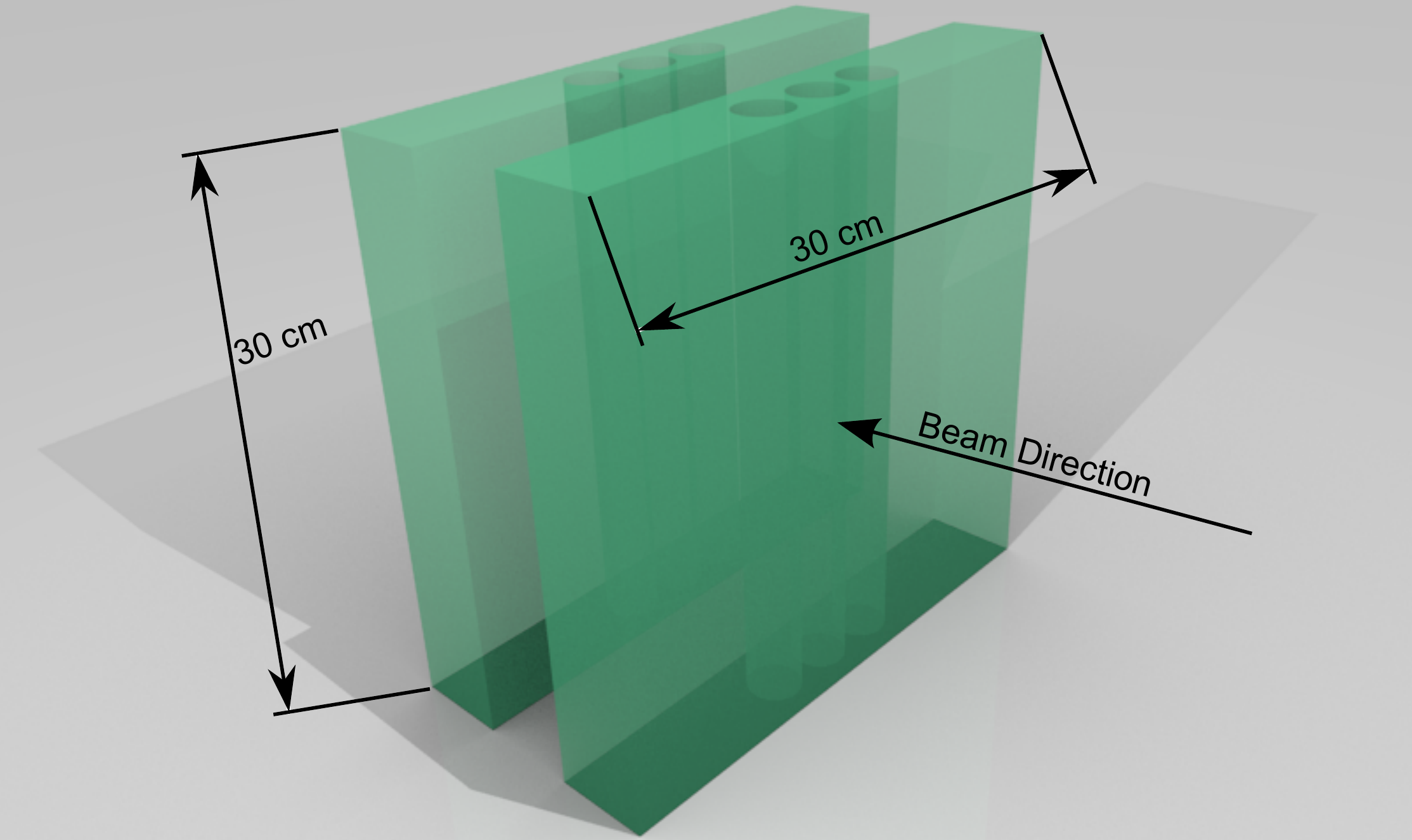}}
    & \subfloat[ ] {\label{subfig:BeamProfile} \includegraphics[width=0.49\textwidth]{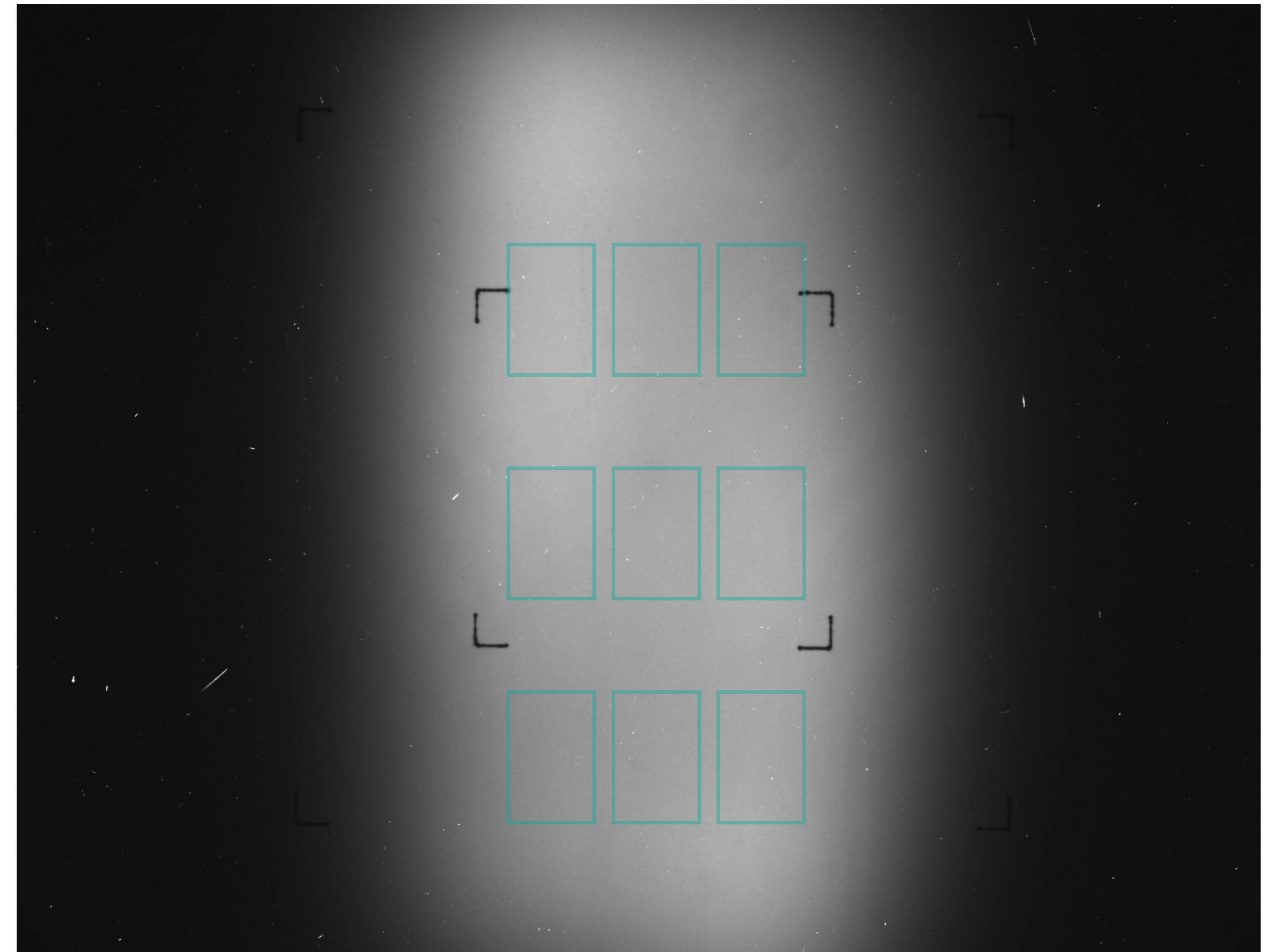}}   \\
\end{tabular}
\label{fig:IrradDetails}
\caption{Details of the proton beam irradiation geometry; (a) 
a schematic drawing of the sample holders used in the irradiations, (b) 
the proton beam intensity profile incident upon the scintillator volumes (outlined areas), taken using a phosphor screen coupled to a CCD. The fiducial marks in (b) 
define the 10 cm by 10 cm region about the beam centre.}
\end{figure*}

\setlength{\tabcolsep}{11pt}

\section{Experiment}

\subsection{Materials}
For the pure liquid scintillator samples we used a Linear Alkyl Benzene (LAB) solvent with 3 g/L 2,5-Diphenyloxazole (PPO) as primary fluorophore and 15 mg/L 1,4-Bis(2-methylstyryl)benzene as a wavelength shifting fluorophore. Our WbLS was composed of 5\% LAB by mass with 2 g/L PPO in the organic phase as the primary fluorophore.

\subsection{Irradiation A}
Proton irradiations were carried out at the NASA Space Radiation Laboratory (NSRL) at Brookhaven National Laboratory. Duplicate samples in polypropylene vials were mounted in two HDPE mounts and exposed to a 10 cm x 20 cm beam of 201 MeV protons (figure \ref{subfig:BeamSetup}).
The uniformity of the beam intensity across the target area was measured to be approximately $\pm 13 \%$ (figure \ref{subfig:BeamProfile}).
Two consecutive beam exposures were used to deliver approximately 300 Gy and a further 500 Gy to samples of pure liquid scintillator and WbLS. Samples from the rear mount were removed following the first exposure, and samples in the front mount were irradiated by both exposures.
The dose accumulated in the rear mount was verified using a Fricke dosimetry solution that was placed in the sample holder along with the scintillating samples. The Fricke measurements suggest that the first exposure delivered a dose of 328 $\pm$ 10 Gy to the Fricke vial. The 800 Gy accumulated by the front holder's samples is twice the maximum dose measurable with Fricke dosimetry \cite{ASTMFricke}, so the NSRL beam instrumentation -- a calibrated ionisation chamber placed upstream from the samples -- was relied upon for the dose estimate. 

During the irradiations, online measurements of the scintillation response of liquid scintillator and WbLS samples were taken to search for short term radiation damage effects. Liquid Scintillator and WbLS samples were placed 50 cm behind the HDPE holders and measurements were carried out using photomultipliers and a digital oscilloscope. A gas ionisation chamber was used as a reference to correct for variations in the beam intensity. No short term damage to the light yield of either sample was observed, to a precision of 3.3\% and 2.4\% for the WbLS and liquid scintillator samples, respectively. The dose delivered to the samples is estimated to be 556 Gy, using the simulation model described below.

Empty vials were also placed in the beam area and later filled with scintillator and water-based liquid scintillator to check for radiation-induced leaching from these vials. Measurements of the scintillating materials that resided in these vials for approximately 1 month revealed no change in the light yield, suggesting that vial leaching does not affect our results.

\begin{table*}[bht]
\renewcommand{\arraystretch}{1.3}
\caption{The simulated mean energy deposit per interacting proton, relative beam intensity, and relative dose in the irradiated samples. All results refer to a vial placed at the beam centre.}
\label{tab:DoseInfo}
\centering
\begin{tabular}{cccc}
\hline
\hline
Sample Location & Mean Energy Deposit   & Relative beam intensity   & Relative dose          \\
\hline
Front HDPE holder       & 9.37 MeV              & 1                         & 1                      \\
Rear HDPE holder        & 9.75 MeV              & 0.957                     & 0.996                  \\
Online measurement      & 9.99 MeV              & 0.652                     & 0.695                 \\
\hline
\hline
\end{tabular}
\end{table*}

\subsubsection{Simulation Model}
\begin{figure}[tbh]
\centering
\includegraphics[width=\columnwidth]{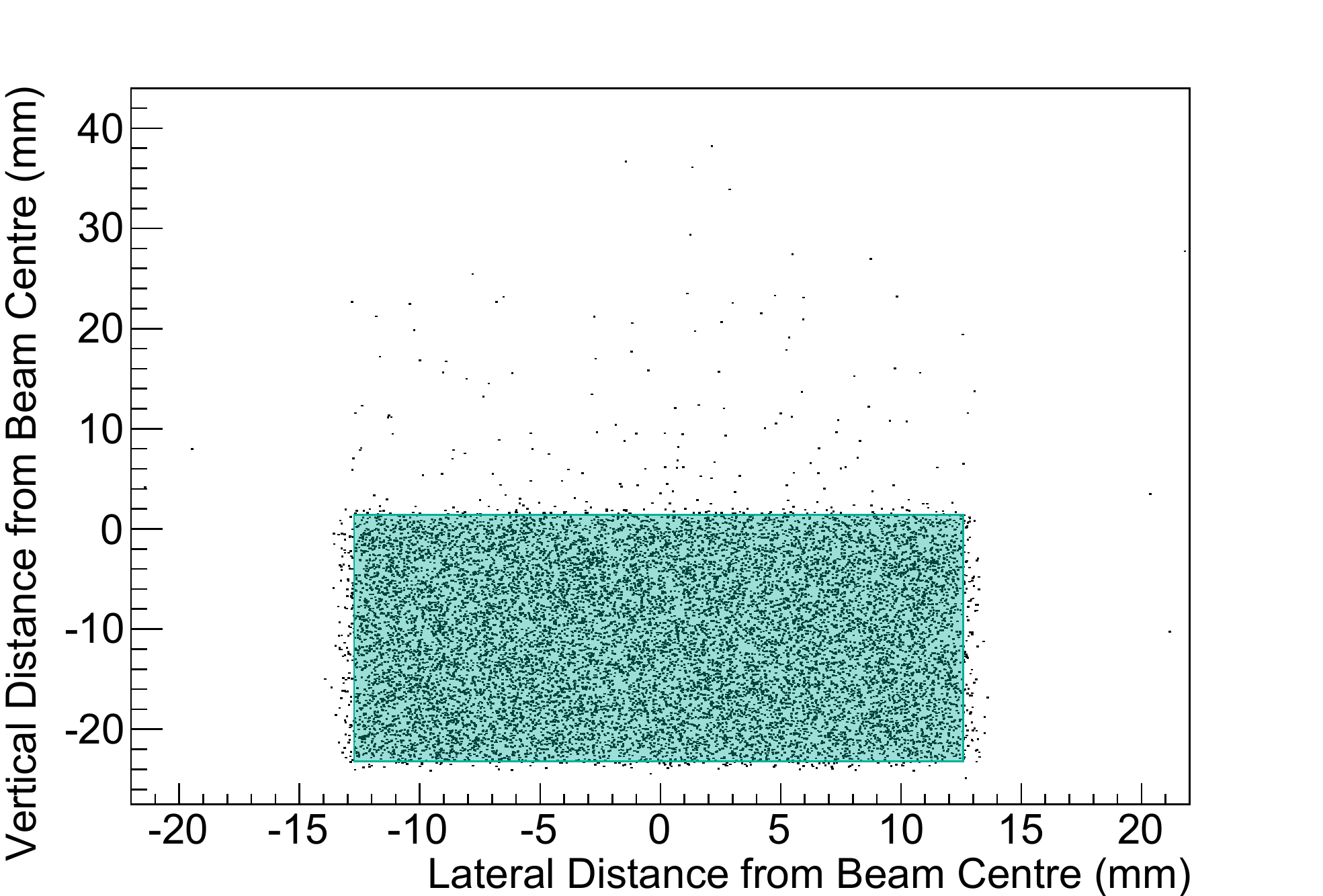}
\caption{The initial beam location of simulated protons that deposit energy in the scintillator encapsulated by the central rear vial. The shaded region denotes the location of the scintillator. }
\label{fig:Verticies}
\end{figure}

We have modeled our irradiation experiments using a Geant4 v10.0 \cite{Agostinelli2003etal} simulation of the measurements. The calculated dose information for the irradiated samples is presented in table \ref{tab:DoseInfo}. The doses delivered to the front and rear vials were identical to within 0.5\%. Figure \ref{fig:Verticies} plots the original beam position of protons that interact in the rear vial. The relatively sharp image of the liquid volume suggests the beam divergence upon entry into the rear vial is on the order of 1 mm. Since the beam spreading was small compared to the length scale of the variations in the incident beam intensity (figure \ref{subfig:BeamProfile}), the incident beam profile was used to estimate the relative dose at the sample locations for the rear sample holder.  We attribute the excess events above the scintillating volume in figure \ref{fig:Verticies} to delta rays that travel through the air above the scintillator.

\subsection{Irradiation B}
A second irradiation was carried out to investigate the effect of low doses upon the scintillator light yield. A beam of 201 MeV protons was used to deliver a dose of approximately 20 Gy and 50 Gy in two irradiations to an irradiation geometry similar to figure \ref{subfig:BeamSetup}. A 21.5 cm thick HDPE block was used to slow the incident proton beam, allowing the vials in the rear holder to receive a greater dose from stopping protons. Fricke dosimetry measurements of the front and rear vials suggest that the dose ratio in the rear relative to the front was 1.6. A beam profile similar to figure \ref{subfig:BeamProfile} was obtained.

\subsection{Light yield measurements}
\begin{figure}[!t]
\centering
\includegraphics[width=\columnwidth]{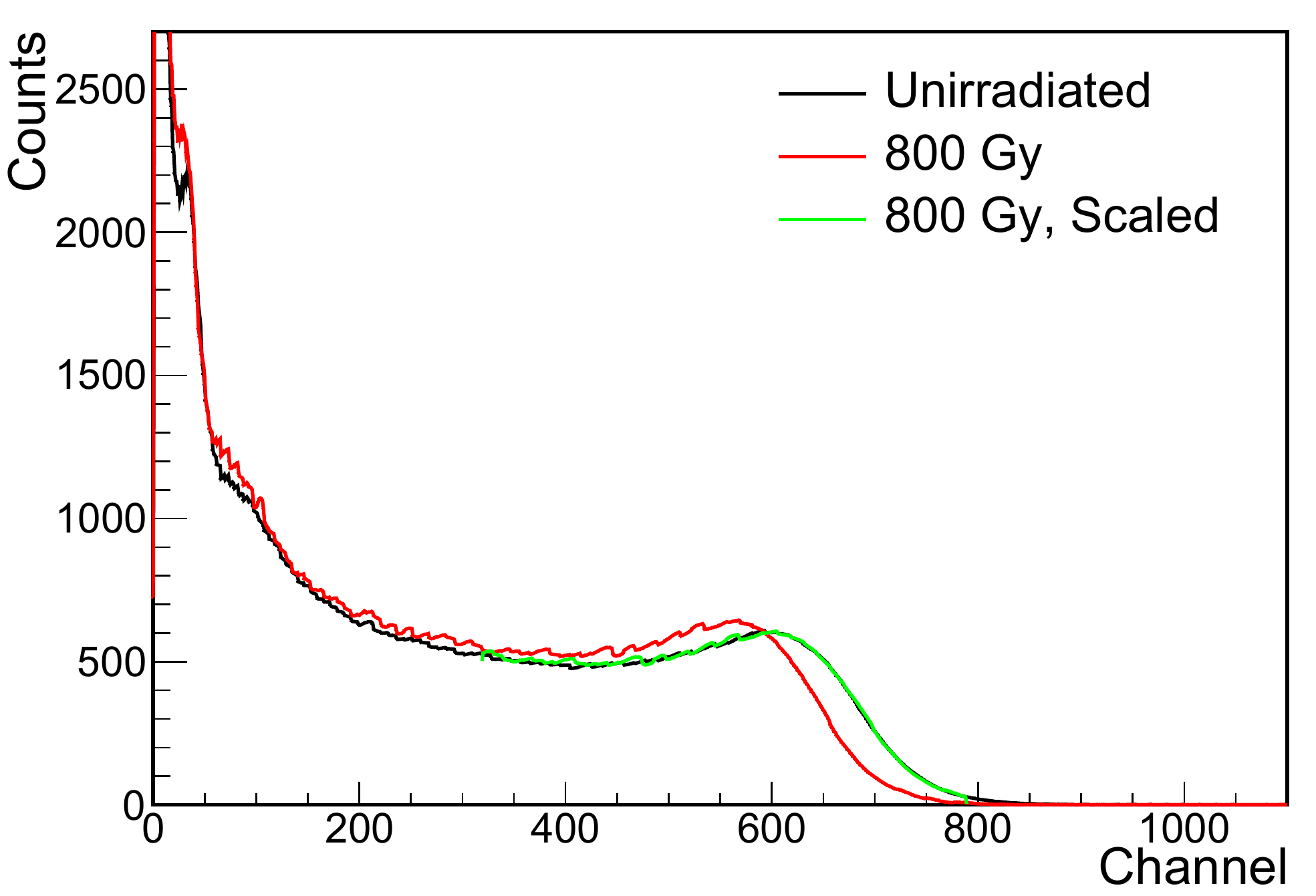}
\caption{An example of the measured Compton electron spectrum for the control (black) and sample of liquid scintillator irradiated by $\approx$800 Gy of 201 MeV protons (red). The green trace shows the irradiated spectrum scaled by the amount of light yield loss as determined by the algorithm described in the text. The non-statistical fluctuations are an artifact of the liquid scintillation spectrometer readout.}
\label{fig:Spec}
\end{figure}

Scintillation measurements were carried out using a Beckman LS6500 liquid scintillation spectrometer. A 
$^{137}$Cs source was placed inside the instrument's lead shielding near the sample chamber to excite the scintillator. The Compton electron spectrum due to this source was measured repeatedly for all samples both prior to and after the irradiation. All irradiated scintillating samples were dispensed into unirradiated vials and compared with an unirradiated control. 
Some instrumental artifacts were observed in the measured scintillation spectrum. These fluctuations in the histogrammed pulse height data exhibited a bin-to-bin correlation in a quasi-periodic fashion (figure \ref{fig:Spec}). The amplitude of these systematic variations did not vary strongly with the number of events in the histogram, so that acquiring for longer periods could minimise this effect. The precise cause of these artifacts is not known. The effect of the instrumental artifacts is taken into account in the assigned systematic uncertainty.

The change in the scintillation light yield upon irradiation was assessed relative to the control sample using an algorithm that compared the sample's Compton electron spectrum with that of the control.
A multiplicative factor was used to scale the sample spectrum along the channel axis. The histogram binning was correctly accounted for in the scaling operation. The change in the measured light yield was determined as the scaling factor that minimised the $\chi ^{2}$ between the measured and control spectra in a region above a channel threshold and below the bin in the control spectrum where the number of counts fell below 20. The optimal channel threshold was determined empirically for each sample. The operation of this algorithm is illustrated in figure \ref{fig:Spec}.

To account for the possibility that our algorithm's multiplicative factor is a biased estimator of the light yield, a Geant4 v10.0 simulation was developed to synthesize a similar Compton electron spectrum to that measured in our detector.
An isotropic source of 662 keV gamma rays was placed 2.5 cm from a scintillating volume identical in size to that used in our measurements. A 2.5 cm thick cylindrical lead shield  with an inner diameter of 7.5 cm surrounded the scintillating volume and source.
The simulated spectrum of energy deposits in the scintillator was convolved with a Gaussian of sufficient width to give an empirical match for the measured liquid scintillator and WbLS spectra.
Systematic errors in our light yield algorithm were investigated by comparing the algorithm's estimate of the scaling factor with the known scaling factor. This analysis shows that the algorithm underestimates the amount of light yield loss for both the pure liquid scintillator and the WbLS, with the lower amplitude WbLS spectrum having a larger error. The magnitude of this error was also found to increase as the reduction in the light yield increased. The required correction was less than 0.5\% of the light yield for all measured light yield losses, and we have corrected our reported results for this effect.

\setlength{\tabcolsep}{0pt}
\begin{figure*}[bt]
\centering
\begin{tabular}{cc}
    \subfloat[Liquid Scintillator] {\label{subfig:PureLSAbs} \includegraphics[width=0.49\textwidth,clip=true,trim=0.5cm 6cm 1.15cm 6cm]{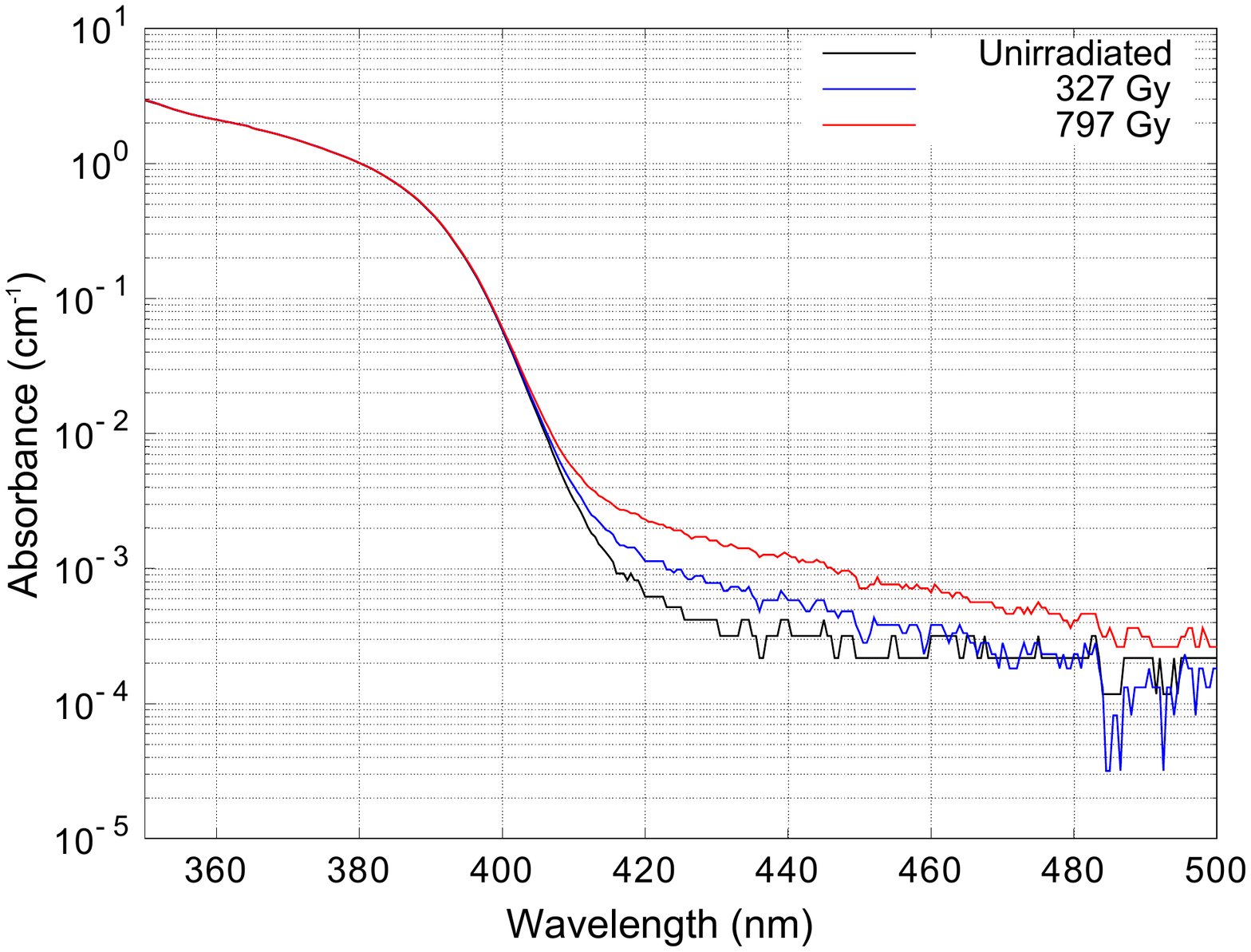}}
    & \subfloat[Water-based Liquid Scintillator] {\label{subfig:WbLSAbs} \includegraphics[width=0.49\textwidth,clip=true,trim=0.5cm 6cm 1.15cm 6cm]{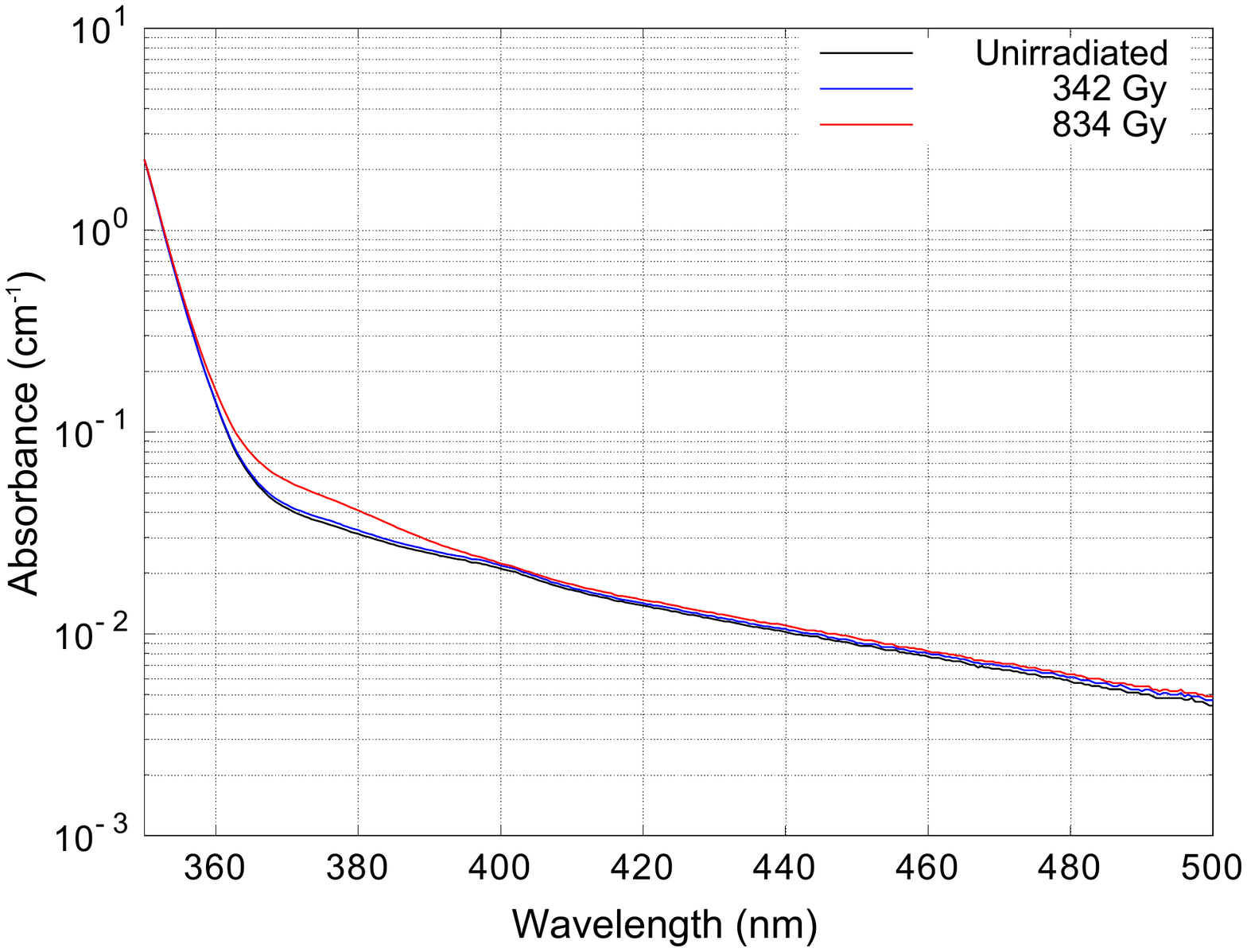}}   \\
\end{tabular}
\caption{The optical absorbance spectra for unirradiated samples (black), and samples irradiated with protons (blue and red). The average doses received by samples in the same irradiation holder are given.}
\label{fig:AbsSpec}
\end{figure*}
\setlength{\tabcolsep}{11pt}

\subsection{Optical Characterisation}
Optical absorbance measurements of the irradiated samples were carried out using a Shimadzu UV1800 spectrophotometer. We observed an increase in the optical attenuation coefficient of the irradiated scintillators that increased with increasing dose delivered to the scintillator (figure \ref{fig:AbsSpec}).

\section{Results}
\begin{figure}[tbh]
\centering
\includegraphics[width=\columnwidth,clip=true,trim=0.5cm 6cm 0.5cm 6cm]{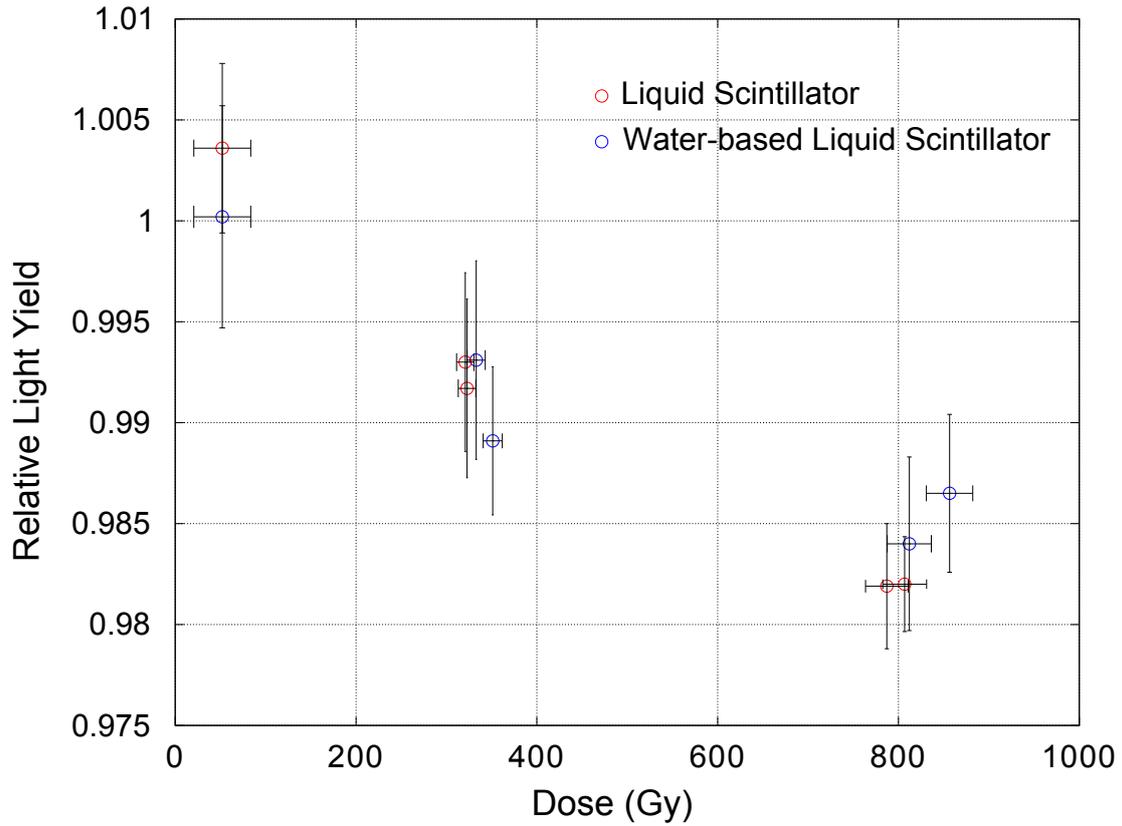}
\caption{The measured light yield in the liquid scintillator (red) and Water-based Liquid Scintillator (blue) following proton irradiation, relative to an unirradiated control.}
\label{fig:LightYields}
\end{figure}

The reduction of the light yield at the proton doses received by the samples is given in figure \ref{fig:LightYields}. The samples were measured approximately 24 hours following irradiation to allow short lived activation products to decay. No increase in the count rate was observed relative to the pre-irradiation measurements, suggesting that long lived activation products were created at negligible levels. All samples from Irradiation B exhibited no significant change in the light yield relative to the control samples. Their average dose and relative light yield is given in the figure, with the dose uncertainty covering the range of doses received by these samples (20 Gy to 90 Gy). The light yield uncertainty for all samples is the combination of the measurement standard deviation 
and an estimated relative uncertainty on the bias correction of 50\%. A small loss of light yield was observed at approximately 320 Gy for both WbLS and liquid scintillator samples. For doses of approximately 800 Gy, a clear degradation of the scintillation yield is evident in both samples. This damage level agrees with the online measurements within their uncertainty, suggesting that there was no measurable recovery of scintillator in the time between the irradiation and the measurement. The scintillation light yield was measured repeatedly for approximately one week for all samples following irradiation, and no recovery or further deterioration was observed.

The optical absorbance spectra (figure \ref{fig:AbsSpec}) of both samples exhibit an increase with dose at wavelength regions relevant for the detection of scintillation photons. The reduction in the measured light yield of this increased optical absorption can be quantified using an effective absorbance due to the radiation damage, $\alpha_{eff}$: 
\begin{equation}
f_{LY} = e^{-\bar{x}\alpha_{eff}}
\end{equation}
where $f_{LY}$ is the factor by which the light yield is reduced, and $\bar{x}$ is the mean optical path length of the emitted photons within the scintillator. The effective absorbance is calculated by weighting the excess damage-induced absorbance ($\alpha(\lambda)$, the difference between the irradiated and unirradiated traces in figure \ref{fig:AbsSpec}) with the normalised scintillator emission spectrum ($S(\lambda)$) and photomultiplier quantum efficiency ($QE(\lambda)$):
\begin{equation}
\alpha_{eff} = \int \alpha(\lambda)S(\lambda)QE(\lambda) d\lambda
\end{equation}
The amount of the observed light yield change that is due to the damage-induced optical absorbance is given in table \ref{tab:OptAbsEffect}, approximating $\bar{x}$ to the radius of the sample vial. The additional optical absorbance induced in the samples is insufficient to fully account for the observed change in the scintillation light yield, which suggests that the damage is also evident as a reduction in the scintillation photon production. This may be due to the radiation-induced formation of chemical quenching agents, or modification of the solvent/fluorophore molecules involved in the scintillation process \cite{Chong1997, Mesquita2001}. We performed Fourier transform infrared spectroscopy to identify chemical changes in the samples. No significant changes were observed. Given that the doses in Refs \cite{Chong1997, Mesquita2001} were orders of magnitude larger than the maximum dose of this work, this is perhaps unsurprising.

\begin{table*}[tbh]
\renewcommand{\arraystretch}{1.3}
\caption{The effect of damage-induced optical absorbance upon the measured light yield.}
\label{tab:OptAbsEffect}
\centering
\begin{tabular}{x{3.4cm}cx{2.5cm}x{4.5cm}}
\hline
\hline
Material                        & Mean Dose          & Measured light yield loss & Light yield loss expected from optical absorbance \\
\hline
Liquid Scintillator             & 327 $\pm$ 10 Gy    & 0.7 $\pm$ 0.4 \%        & 0.1 $\pm$ 0.2\%                                            \\
Liquid Scintillator             & 797 $\pm$ 24 Gy    & 1.7 $\pm$ 0.3 \%        & 0.3 $\pm$ 0.2\%                                            \\
Water-based Liquid Scintillator & 342 $\pm$ 10 Gy    & 0.8 $\pm$ 0.5 \%        & 0.0 $\pm$ 0.2\%                                           \\
Water-based Liquid Scintillator & 834 $\pm$ 25 Gy    & 1.3 $\pm$ 0.5 \%        & 0.7 $\pm$ 0.2\%                                            \\
\hline
\hline
\end{tabular}
\end{table*}

\section{Discussion and Conclusion}
No damage to the scintillation light yield was evident for proton doses up to 90 Gy. Given that a typical patient dose in a proton therapy treatment fraction is 2 Gy, this suggests that a liquid scintillator-based proton therapy dose-verification system could service at least 45 patients with no change in performance.

We will now consider the performance of such a QA instrument in more detail, given the observed changes at 800 Gy of proton dose. The typical annual number of patients served by a single proton therapy treatment room is approximately 300. Therefore, a QA instrument that validates the treatment plan of every patient may receive approximately 600 Gy in a year of operation, assuming that the instrument is used to validate a single 2 Gy treatment fraction for each patient. While our study has shown that the scintillator is damaged by this level of dose; an important consideration is that the entire volume of the scintillating phantom is not irradiated during the QA measurement. Indeed, for a phantom volume of 30$^{3}$ cm$^{3}$, and a typical lateral target size of 9 cm$^{2}$ and target depth of 20 cm, the irradiated volume is approximately 1\% of the total phantom volume. We assume a flat dose profile delivering uniform dose to this volume, which is a conservative approximation. Therefore, for complete mixing of the irradiated scintillator with the larger volume, the light yield damage is diluted by a factor of 100 relative to our measurements. In our measurement geometry, 600 Gy of proton dose to a scintillator results in an apparent light yield loss of approximately 1.3\%. Half of this loss is assumed to be due to damage-induced optical attenuation increases, which corresponds to $\alpha_{eff} = 0.0083$ cm$^{-1}$. Therefore, the irradiated phantom volume would exhibit a scintillation light production reduction of $ 1.3\% / (2 \times 100) = 0.007\%$. For a mean optical path length of 15 cm, the light absorption in the scintillator volume would decrease the detected number of photons by $1 - e^{-15*0.0083/100} = 0.1\%$. Therefore, the amount of light yield reduction due to radiation damage in a QA phantom will be approximately 0.1\% over a year of full-time clinical use. It is probable that this level of systematic variation of a dose verification instrument is acceptable for clinical operation given the magnitude of other uncertainties in the dose measurement \cite{ICRU78ch8}.

\begin{acknowledgments}
The authors would like to thank Mike Sivertz, Adam Rusek, and Chiara La Tessa at the NASA Space Radiation Laboratory, as well as Russ Burns, Rich Sautkulis, and Jim Jardine for their assistance with this study.
This research was funded by a Technology Maturation Award from the Office of Technology Commercialization and Partnerships at Brookhaven National Laboratory.
\end{acknowledgments}

\bibliographystyle{unsrt}
\bibliography{library.bib}

\end{document}